\documentclass[fleqn, useAMS,usenatbib]{mn2e}

\usepackage{ amssymb, graphicx, verbatim}
\usepackage{ upgreek, amsmath}

\newcommand{\e}[1]{\times 10^{#1}}

\newcommand{\wl}{$\lambda$}
\newcommand{\msun}{M$_\odot$}
\newcommand{\wll}{$\lambda \lambda$}

\def\ni {$^{56}$Ni}
\def\nif {$^{58}$Ni}
\def\co {$^{56}$Co}

\def\kms {km~s$^{-1}$}
\def\ergs {erg s$^{-1}$}
\setcitestyle{round,aysep={},yysep={,}}
\makeatletter
\def\input@path{{../matlab/}{../figs/}}
\makeatother

\title[SN 2012ec]{Supersolar Ni/Fe production in the Type IIP SN 2012ec}
\author[Anders Jerkstrand]{A. Jerkstrand$^{1}$\thanks{E-mail:a.jerkstrand@qub.ac.uk}, 
S. J. Smartt$^{1}$, J. Sollerman$^{2}$, C. Inserra$^{1}$, M. Fraser$^{3}$,  J. Spyromilio$^4$,
\newauthor C. Fransson$^{2}$, T.-W. Chen$^1$, C. Barbarino$^{5,6}$, M. Dall'Ora$^6$, M. T. Botticella$^6$,
\newauthor  M. Della Valle$^{6,7}$, A. Gal-Yam$^8$, S. Valenti$^{9,10}$, K. Maguire$^4$, P. Mazzali$^{11,12,13}$,
\newauthor L. Tomasella$^{12}$\\
$^1$Astrophysics Research Centre, School of Mathematics and Physics, Queen's University Belfast, Belfast BT7 1NN, UK
\\
$^2$The Oskar Klein Centre, Department of Astronomy, Stockholm University, Albanova, 10691 Stockholm, Sweden\\
$^3$Institute of Astronomy, University of Cambridge, Madingley Road, Cambridge CB3 0HA, UK \\
$^4$ESO, Karl-Schwarzschild-Strasse 2, 85748 Garching, Germany \\
$^5$Dip. di Fisica and ICRA, Sapienza Universit\`a di Roma, Piazzale Aldo Moro, I-00185 Rome, Italy\\
$^6$INAF - Osservatorio Astronomico di Capodimonte, Salita Moiariello 16, 80131 Naples, Italy\\
$^7$ICRANet-Pescara, Piazza della Repubblica 10, I-65122 Pescara, Italy\\
$^8$Department of Particle Physics and Astrophysics, Weizmann Institute of Science, Rehovot 76100, Israel\\
$^9$Las Cumbres Observatory Global Telescope Network, 6740 Cortona Dr., Suite 102, Goleta, CA 93117, USA\\
$^{10}$Department of Physics, University of California, Santa Barbara, Broida Hall, Mail Code 9530, Santa Barbara, CA 93106-9530, USA\\
 Karl-Schwarzschild-Str. 2, D-85748 Garching, Germany\\
$^{11}$Astrophysics Research Institute, Liverpool John Moores University, 146 Brownlow Hill, Liverpool L3 5RF, UK\\
$^{12}$INAF - Osservatorio Astronomico di Padova, I-35122 Padova, Italy\\
$^{13}$Max-Planck Institut f\"ur Astrophysik, Karl-Schwarzschildstr. 1, D-85748 Garching, Germany
}
\begin{document}

\date{}

\pagerange{\pageref{firstpage}--\pageref{lastpage}} \pubyear{2002}

\maketitle

\label{firstpage}

\begin{abstract}
SN 2012ec is a Type IIP supernova (SN) with a progenitor detection and comprehensive photospheric-phase observational coverage. Here, we present Very Large Telescope and PESSTO observations of this SN in the nebular phase. We model the nebular [O I] \wll6300, 6364 lines and find their strength to suggest a progenitor main-sequence mass of $13-15$ \msun. SN 2012ec is unique among hydrogen-rich SNe in showing a distinct line of stable nickel [Ni II] \wl7378. This line is produced by \nif, a nuclear burning ash whose abundance is a sensitive tracer of explosive burning conditions. Using spectral synthesis modelling, we use the relative strengths of [Ni II] \wl7378 and [Fe II] \wl7155 (the progenitor of which is \ni) to derive a Ni/Fe~production ratio of $0.20\pm0.07$ (by mass), which is a factor $3.4\pm1.2$ times the solar value. High production of stable nickel is confirmed by a strong [Ni II] 1.939 $\mu$m line. This is the third reported case of a core-collapse supernova producing a Ni/Fe ratio far above the solar value, which has implications for core-collapse explosion theory and galactic chemical evolution models.
\end{abstract}

\begin{keywords}
supernovae: general - supernovae: individual: SN 2012ec - stars: evolution 
\end{keywords}
\section{Introduction}
Research on the elusive core-collapse supernova (SN) explosion mechanism is ongoing in multiple groups around the world \citep[see][for a review]{Janka2012}. As the core of the star collapses to a neutron star, a shock wave is born that travels out through the infall of the mantle. The initial energy of the prompt shock is not enough to reverse the accretion, but a delayed explosion can be obtained if neutrino heating is efficient enough. Several successful explosions have been obtained in recent simulations with self-consistently calculated neutrino luminosities, e.g. by \citet{Kitaura2006} in spherically symmetric models of electron-capture supernovae, and with additional support from multi-dimensional instabilities in more massive progenitors \citep{Marek2009, Muller2012, Bruenn2013}. 

Important constraints on the explosive process can be obtained by observations and modelling of SNe in the  nebular phase, when the ejecta become optically thin and the inner regions of nucleosynthesis products become visible. The deepest layers of iron-group nuclei, which are the ashes of explosive silicon burning, are directly associated with the critical gain layers where the neutrinos power the shock wave, and are therefore direct diagnostics of the explosion process.


Type IIP (Plateau) SNe are the most common core-collapse explosions \citep[$\sim50\%$ per unit volume,][]{Li2011}. Light curve models \citep{Chevalier1976} and progenitor detections \citep{Smartt2009} have shown that these are explosions of red supergiant (RSG) stars. Progenitor luminosities \citep{Smartt2009} and nucleosynthesis yields \citep[][J14 hereafter]{Jerkstrand2014} suggest helium core masses $M_{\rm He} \lesssim 5$ \msun~($M_{\rm ZAMS} \lesssim 18$ \msun) for the progenitor population, whereas hydrodynamical modelling favours a more extended mass range \citep[e.g.][]{Utrobin2009, Dallora2014}\footnote{The light curve is most sensitive to the hydrogen envelope mass (see discussion in \citet{Dessart2013}), with best-fitting values being 20-30 \msun~for some SNe, favouring high-mass progenitors.}. 

SN 2012ec is a Type IIP explosion that occurred in NGC 1084 in early August 2012. A potential progenitor was reported by \citet{Maund2013}, a high luminosity star with $\log{\left(L/L_\odot\right)} = 5.0-5.4$. Stars with luminosity on the upper end of this range have not yet been seen to explode as RSGs \citep{Smartt2009}. 
The possibility that the progenitor was a luminous, high-mass star, motivated us to embark on a follow-up campaign to follow the SN through its photospheric and nebular phases. The photospheric-phase PESSTO (Public ESO Spectroscopic Survey of Transient Objects) data is presented by Barbarino et al. (2014, submitted). Here we report on PESSTO and VLT (Very Large Telescope) observations in the nebular phase, and modelling of these data. 
For the analysis, we follow \citet{Maund2013} and assume an explosion epoch of August 5 2012, a distance of 17.3 Mpc, an extinction $E_{B-V} = 0.11$ mag , and a heliocentric recession velocity of 1407 \kms \citep{Koribalski2004}. All observed spectra displayed in the paper have been dereddened and redshift corrected.


\section{Observations and data reduction}
\label{sec:obs}

\subsection{Photometry}
\label{sec:photdata}


Optical and near-infrared imaging was obtained for the phase $+176$ to $+551$ days after explosion using the European Southern Observatory's New Technology Telescope (NTT) with EFOSC2 and SOFI, the SMARTS 1.3m telescope (operated by the SMARTS consortium) using ANDYCAM, and the Liverpool 2m Telescope (LT) using RATCam. Tables \ref{table:opticalphot} and \ref{table:nirphot} present the photometry. The $V$, $R$, $I$ magnitudes are template-subtracted (see below) whereas $B$, $r$, $z$ are not (as no templates were available in these filters). The NTT data were collected as part of the PESSTO program \citep{2013Msngr}, and are a continuation of the public monitoring campaign of SN 2012ec described in \citet{Maund2013} and Barbarino et al. (2014, submitted). 
The EFOSC2 images were reduced (trimmed, bias subtracted, and flat-fielded) using the PESSTO pipeline as described in Smartt et al. (2014, submitted), whereas the LT and SMARTS images were reduced automatically using their respective pipelines.

Photometric zero-points and colour terms were computed through observations of Landolt standard fields \citep{Landolt1992}. Three of the seven nights in which optical imaging was carried out were photometric and we calibrated the magnitudes of a local stellar sequence using these data. We chose the reference stars 1, 11, and 12 presented in Barbarino et al. (2014, submitted), and found reasonable agreement in the computed magnitudes of the secondary standards (the differences were within $0.05$ mag). The average magnitudes of the local-sequence stars were used to calibrate the photometric zero-points obtained in non-photometric nights, or when the colour terms were not retrieved.

The complex and high background galaxy flux at the position of SN~2012ec meant that image template subtraction was required on all the EFOSC2 $VRI$\footnote{EFOSC2 mounts the Gunn $i$ \#705 filter that we calibrated to Johnson-Cousin $I$.} data. Fortunately the large  programme 184.D-1140 (PI: S. Benetti) had observed SN 2009H in the same galaxy and $VRi$ images from 10 October 2009 with EFOSC2 were available to construct pre-discovery templates. Each filter had 5$\times$120 second frames taken, which were co-added and then subtracted from the target frames using the HOTPANTS\footnote{http://www.astro.washington.edu/users/becker/hotpants.html} image subtraction software \citep[which is based on the algorithm presented in][]{Alard2000}. 

 For the near-infrared (NIR) observations, we took multiple, dithered, on-source exposures; these images were then flat-fielded and median-combined to create a sky frame. The sky frame was subtracted from each of the individual images, which were then aligned and co-added. The total exposure times in the two epochs which had detections were 400s ($J$), 360s ($H$) and 1080s ($K_{s}$) on 28 January 2013 and 400s ($J$), 360s ($H$) and 720s ($K_{s}$) on 21 February 2013. NIR photometry of the reference stars was calibrated using the Two Micron All Sky Survey (2MASS) catalogue magnitudes \citep{Skrutskie2006}. The full reduction procedure is described in Smartt et al. (2014, submitted) and these reduced images have been released as part of the PESSTO SSDR1. Users should note that the on-source dithering pattern results in the over-subtraction of the sky background in some of the images, but the  photometry of the SN is not affected. Further imaging was taken a year later on 6 February 2014 with exposure times of 200s ($J$), 150s ($H$) and 900s ($K_{s}$). SN 2012ec was not detected on these images. We used these images as templates so that the values at 177 and 201 days in Table \ref{table:nirphot} are template subtracted. 


Photometric flux measurements were performed using a point-spread function (PSF) fitting technique. We  simultaneously fitted the PSF of SN 2012ec and the sequence stars using the {\sc snoopy}\footnote{based on the {\sc iraf daophot} package.} package within {\sc iraf}\footnote{{\sc iraf} is distributed by the National Optical Astronomy Observatory, which is operated by the Association of Universities for Research in Astronomy (AURA) under cooperative agreement with the National Science Foundation.}. 

\begin{table*}
	\begin{center}
	\caption{Optical photometry. $V$, $R$, $I$ magnitudes are template-subtracted whereas $B$, $r$, $z$ are not. $B$, $V$, $R$, $I$ magnitudes are in the Landolt Vega system, whereas $r$, $z$ are in the AB system. The phase is relative to the estimated explosion epoch of August 5 2012 (MJD 56144).}
	\label{table:opticalphot}
		\begin{tabular}{ccccccccc}
		
			\hline 
		 	    Date   & Phase   & Telescope  & $B$        & $V$            & $R$            & $r$            & $I$            & $z$\\
						\hline
		 	08/02/2013  & +188    & SMARTS & $19.29\pm0.09$ & $18.47\pm0.08$ & $17.29\pm0.08$ & ...            & $16.88\pm0.08$ & ...\\
			23/02/2013  & +202    & LT     & $19.31\pm0.10$ & $18.62\pm0.09$ & ...            & $17.34\pm0.15$ & $17.19\pm0.09$ & $17.13\pm0.10$\\
                        01/03/2013  & +208    & SMARTS & ...            & $18.62\pm0.11$ & $17.52\pm0.09$ & ...            & $17.25\pm0.09$ & ...\\
			11/09/2013  & +403    & NTT    & ...            & $20.60\pm0.11$ & $19.19\pm0.09$ & ...            & $20.18\pm0.09$ & ...\\
			07/10/2013  & +429    & NTT    & ...            & $20.78\pm0.10$  & $19.68\pm0.06$ & ...            & $20.38\pm0.09$ & ...\\
			22/11/2013  & +475    & NTT    & ...            & $21.04\pm0.08$ & $20.87\pm0.07$ & ...            & $20.72\pm0.09$ & ...\\
			23/12/2013  & +506    & NTT    & ...            & $22.38\pm0.08$ & $21.53\pm0.11$ & ...            & $21.13\pm0.11$ & ...\\
			31/01/2014  & +545    & NTT    & ...            & $22.66\pm0.15$ & $22.16\pm0.10$  & ...           & ...            & ...\\
				\hline 
\end{tabular}
\end{center}
\end{table*} 


\begin{table*}
	\begin{center}
	\caption{Near-infrared photometry. Magnitudes are in the 2MASS Vega system. The upper limits are 3$\sigma$ values. The first two epochs are template-subtracted.}
	\label{table:nirphot}
		\begin{tabular}{cccccc}
		
			\hline 
			Date & Phase & Telescope & $J$ & $H$ & $K_s$\\
						\hline
			28/01/2013 & +177 & NTT & $16.59\pm0.04$ & $16.18\pm0.04$ & $16.29\pm0.05$\\
			21/02/2013 & +201 & NTT & $17.08\pm0.04$ & $16.55\pm0.04$ & $16.70\pm0.05$\\
		 	06/02/2014 & +551 & NTT & $>20.3$ & $>19.6$ & $>19.8$\\
				\hline 
\end{tabular}
\end{center}
\end{table*} 


\begin{table*}
\caption{Summary of spectroscopic observations.}
\begin{tabular}{cccccccccc}
\hline
Date       & Phase & Telescope & Instrument & Wavelength          & Slit width & Resolution \\
\hline
2013 Feb 7   & +185   & NTT & SOFI + Blue Grism &  $0.95-1.64$ $\mu$m & 1.0\arcsec           & 23 \AA\\ 
2013 Aug 11  & +371   & VLT & X-shooter + UVB/VIS/NIR         &  $0.32-2.45$ $\mu$m & 1.0\arcsec/0.9\arcsec/0.9\arcsec  & 0.9/1.7/2.3 \AA\\ 
2013 Sep 11  & +402   & NTT & EFOSC2 + Grism\#11&  $3380-7250$ \AA      & 1.0\arcsec        & 13 \AA \\      
\hline 
\end{tabular}
\label{table:spectroscopy}
\end{table*}

\subsection{Spectroscopy}
We obtained three spectra in the nebular phase. The first was a near-infrared spectrum at +185 days using NTT and the SOFI near-infrared spectrograph, the second was a combined  optical and NIR spectrum at +371 days using  X-shooter on the ESO Very Large Telescope (via programme 091.D-0608, PI: J. Sollerman). The third was an optical spectrum at +402 days using EFOSC2 on the NTT. The details of the wavelength coverage, grisms employed, and spectral resolutions are listed in Table \ref{table:spectroscopy}. 


The NTT SOFI spectrum was a combination of exposures taken with on-slit nodding to give a total exposure time of 3240s with the Blue Grism (BG) and the GBF order blocking filter. The PESSTO observing sequence and reduction procedure is described in Smartt et al. (2014, submitted). After extraction and wavelength calibration, the spectrum was corrected for telluric absorption with the standard Hip010502 and flux calibrated with the spectrophotometric standard EG274. 
 
The NTT EFOSC2 optical spectrum was reduced using the PESSTO pipeline. Wavelength calibration was performed using the spectra of comparison lamps and adjusted by checking the night sky line positions. The spectrum was corrected for telluric absorption by subtracting a model absorption spectrum and flux calibration was performed using the PESSTO selected spectrophotometric standard stars observed on the same night. 




SN 2012ec was observed with the VLT+X-shooter at an airmass of 1.2 on Aug. 9.3 2013 UT, and again on 
Aug. 11.3 UT at an airmass of 1.1. X-shooter consists of three echelle 
spectrographs, each covering a separate wavelength range (UVB: 300-560 
nm; VIS: 556-1024 nm; NIR: 1024-2480 nm), which are used 
simultaneously through a dichroic. A 1.0\arcsec\ slit was used in the 
UVB arm, and 0.9\arcsec\ slits were used in the VIS and NIR arms. 
Observations were taken while nodding the telescope, moving the SN 
between two positions in the slit to facilitate sky subtraction 
in the NIR arm (using the template {\sc 
XSHOOTER\_slt\_obs\_AutoNodOnSlit}). The spectrum from each arm was 
reduced using version 2.4.0 of the X-shooter 
pipeline\footnote{http://www.eso.org/sci/software/pipelines/} running 
under the Reflex environment. All data were pre-reduced (bias subtracted 
in the UVB and VIS arms, flat fielded, and wavelength calibrated) within 
the pipeline, and the separate echelle orders were merged into a single 
spectrum. The bias level and sky background in the NIR arm was removed 
by subtracting pairs of consecutive spectra taken with the target at 
different locations in the slit. Finally, the spectra were optimally 
extracted, and flux calibrated using spectrophotometric standards.



\section{Spectral synthesis models}
As foundation for our analysis we use the spectral models presented by \citet[][hereafter J12]{Jerkstrand2012} and J14. These models compute the temperature and NLTE excitation/ionization solutions in each zone of the SN ejecta, taking all relevant thermal, non-thermal, and radiative rates into account. Specifically, the modelling consists of the following computational steps: i) Transport and deposition of radioactive decay products (gamma-rays, X-rays, leptons) ii) The distribution of non-thermal electrons created by the radioactivity iii) Thermal equilibrium in each compositional zone iv) NLTE ionization balance for the twenty most common elements v) NLTE excitation structure for about 50 atoms/ions vi) Radiative transfer through the ejecta. The solutions are generally coupled to each other and global convergence is achieved by iteration.

The calculations are applied to the SN ejecta models computed by \citet{Woosley2007}, with some modifications to mimic the effects of multidimensional mixing which is known to be important \citep[e.g.][]{Hammer2010}. Specifically, the metal zones (Fe/He, Si/S, O/Si/S, O/Ne/Mg, and O/C) are macroscopically mixed together with parts of the He/C, He/N, and H zones in a core region between 0 and 1800 km s$^{-1}$. Each zone has an individual filling factor in this core and is distributed over $10^3$ clumps. The model takes dust formation into account by applying a grey absorption coefficient over the core from 250 d, growing with $\Delta \tau_{\rm dust} = 1.8\e{-3}$ d$^{-1}$ (a calibration to the observed dust formation in SN 2004et, J14). Over the period covered here, this dust component has only a small effect on the optical/NIR spectrum as $\tau_{\rm dust}<$ 0.25 up to 400 days. Outside the core reside the remaining helium layers followed by the hydrogen envelope, whose density profile is determined by the one-dimensional hydrodynamic solutions.

In addition to these models, we here computed a $M_{\rm ZAMS}=15$ \msun~model at 370d using a lower $^{56}$Ni mass of 0.03 \msun, as appropriate for SN 2012ec (all models computed in J12 and J14 have $^{56}$Ni masses of 0.062 \msun). The lower \ni~mass was achieved by reducing the mass of the Fe/He zone. The density of this zone was kept the same, and the filling factors of the other core zones were increased by a factor of 1.09 to fill the vacated volume.


\section{Analysis of photometric data}

\label{sec:photometry}

\subsection{The $^{56}$Ni mass}
\label{sec:nimass}
In the early tail phase of a Type II SN ($\sim$150-200d), steady state sets in so the emergent luminosity follows the instantaneous energy input by radioactivity, dominated by the $^{56}\mbox{Ni} \rightarrow ^{56}\mbox{Co} \rightarrow ^{56}$Fe decay chain. At the same time, the escape fraction of gamma rays is neglegible. This phase therefore offers an opportunity to determine the \ni~mass by estimating the bolometric luminosity.

We estimate the $BVRIJHK_s$ quasi-bolometric luminosity at 187 days by combining the observed optical (dereddened) photometry at 187 days with extrapolated (dereddened) NIR photometry from day 176, using an extrapolation  factor $\exp{\left(-11d/111.4d\right)}$. The magnitudes were converted to flux values at the effective wavelengths of the filters and linearly interpolated. Since no useful templates exist in $B$, the $B$ magnitudes reported in Sect. \ref{sec:photdata} are not template-subtracted. To obtain an estimate for the SN $B$ magnitudes, we estimate the contribution from the galaxy by taking the $B-V$ colour from the galaxy model derived in Sect. \ref{sec:galaxymodel} and apply this to the $V$-band template, obtaining an estimate of the galaxy $B$ magnitude. Subtracting this artificial $B$-template from the measured $B$ magnitudes gave estimates for the SN of $B=19.36$ mag at 187d and $B=19.50$ mag at 202d. The resulting $B-V$ colours of the SN are similar to other Type IIP SNe \citep{Maguire2010,Inserra2012,Inserra2013,Tomasella2013}.

We obtain $L_{\rm BVRIJHK_s}^{\rm 187d} = 7.0\e{40}$ \ergs. To estimate the full bolometric luminosity, we assume the fraction outside $B$ to $K_s$ to be the same as in SN 1987A at a similar epoch, which we compute as 19\% at 165 d using data from \citet{Hamuy1988} and \citet{Bouchet1989}, and using $E_{B-V}=0.15$ mag for SN 1987A. We then obtain $L_{\rm bol}^{\rm 187d} = 8.6\e{40}$ \ergs, which from using Eq. 6 in J12 corresponds to an initial \ni~mass of 0.033 \msun. Doing the same evaluation using the optical data at 202 days combined with the NIR data at 200 days gives a \ni~mass of 0.030 \msun. These values are in good agreement with derivations in Barbarino et al. (2014, submitted) based on earlier epochs ($0.04 \pm 0.01$ \msun). Doing the same analysis with $V$ to $K_s$ (to avoid the uncertain $B$ band) gives very similar numbers. For the rest of the analysis in this paper we adopt a \ni~mass of 0.03 \msun, and take the error in this to be 0.01 \msun.

\subsection{Photometric evolution}
\begin{figure*}
\includegraphics[width=1\linewidth]{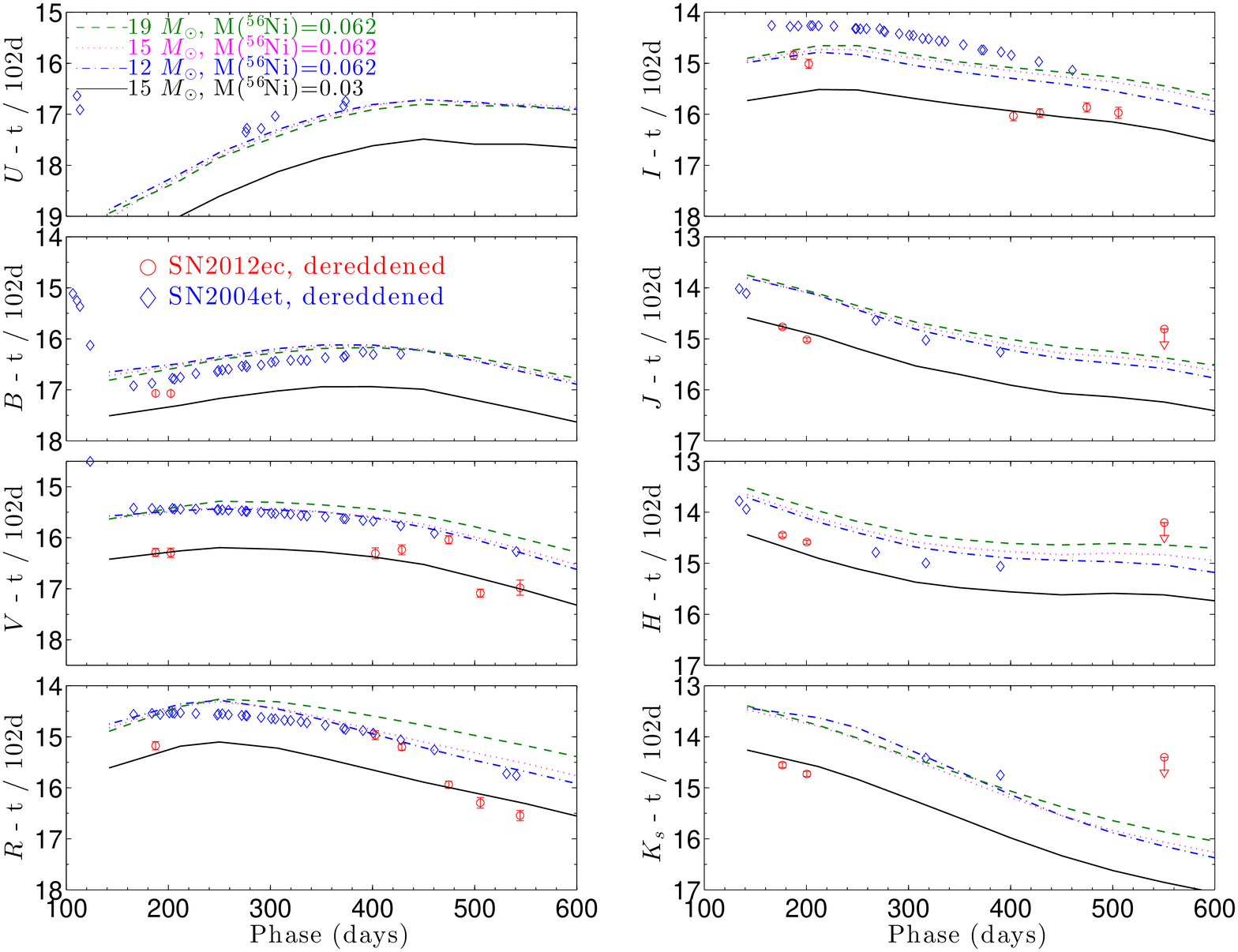} 
\caption{Evolution of $B$ to $K_s$ (subtracted by $t/102d$ to normalize to the $^{56}$Co decay) of SN 2012ec (red points), and SN 2004et \citep[blue points, data from][]{Sahu2006, Maguire2010, Fabbri2011}. The 12, 15, and 19 \msun~models of \citet{Jerkstrand2012}  are plotted as green, magenta, and blue lines, and the 15 \msun\ model scaled with 0.03/0.062 to adjust for a lower \ni~mass of \protect\input{scaledNimass.txt}\msun~is plotted as a black solid line. All observed magnitudes have been corrected for dust extinction and scaled to the distance of SN 2012ec.}
\label{fig:photometry}
\end{figure*}

Figure \ref{fig:photometry} shows the evolution of SN 2012ec (red circles) in $B$ to $K_s$ bands between $176-550$ days post-explosion, compared to the models computed in J12 as well as the observed evolution of the well-studied SN 2004et (blue diamonds). All magnitudes have been normalized to the \co~decay (one mag in 102 days) by having a term $t/102d$ subtracted. The observed magnitudes have been corrected for dust extinction ($E_{B-V}= 0.11$ mag for SN 2012ec and $E_{B-V}=0.41$ mag for SN 2004et) and SN 2004et has been scaled to the same distance as SN 2012ec. 

The early phase photometry (176-202d) shows reasonable agreement with a model scaled with a factor 0.03/0.062 (to adjust for the \ni~mass of 0.03 \msun~in SN 2012ec, see Sect. \ref{sec:nimass}). The data between 400-550d shows quite a lot of scatter. This is likely due to difficulties in the template subtractions, as the background flux is significantly higher than the SN flux at these epochs (Sect. \ref{sec:spectroscopy}). The average flux levels are still in reasonable agreement with the model brightness for a \ni~mass of 0.03 \msun.



\section{Analysis of spectroscopic data}
\label{sec:spectroscopy}

\subsection{Optical spectral modelling}
\label{sec:galaxymodel}
\begin{figure*}
\includegraphics[width=1\linewidth]{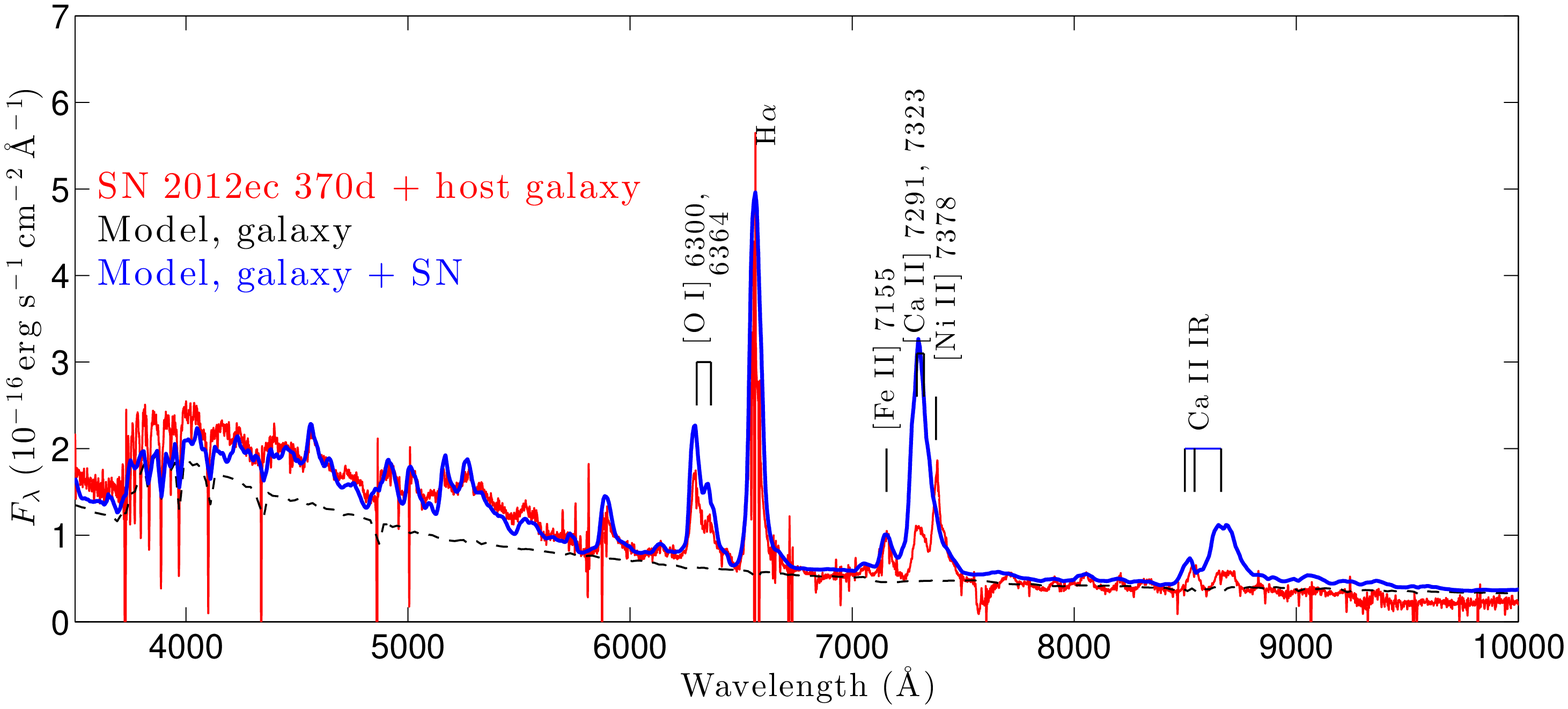} 
\includegraphics[width=1\linewidth]{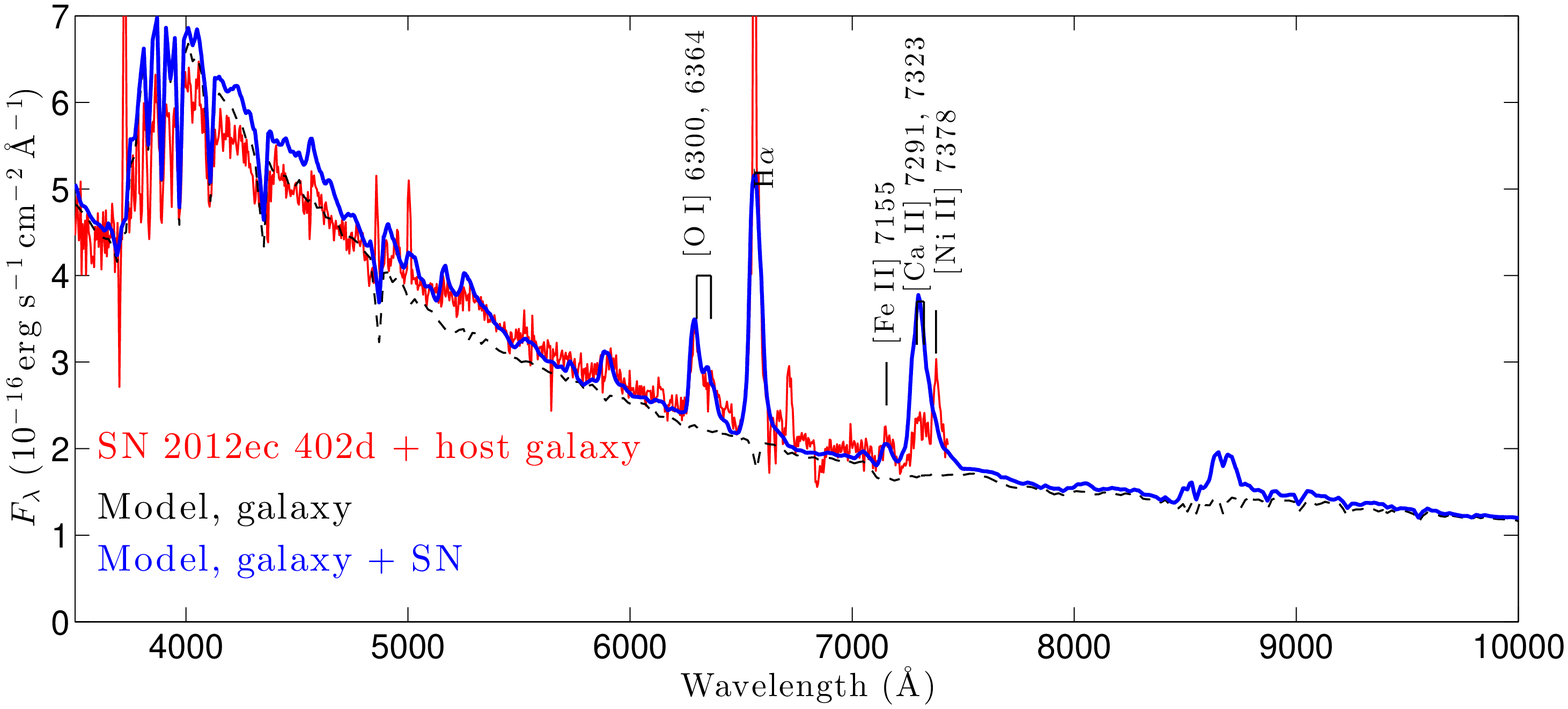} 
\caption{{\bf Top} : Smoothed X-shooter spectrum of the SN 2012ec location at 371 days (red), dereddened and redshift corrected. Also shown is the Starburst99 galaxy model (black dashed), and this galaxy model plus the new 15 \msun\ SN model with a \ni~mass of 0.03 \msun~(blue). {\bf Bottom} : NTT spectrum of SN 2012ec at 402 days (red), dereddened and redshift corrected. Also shown is the same galaxy model as above, but scaled by a factor 3.6 (black dashed), and the sum of this model and the same SN model as above, but scaled with a factor $\exp{\left(-31\mbox{d}/111\mbox{d}\right)}$ to compensate for the later epoch (blue).}
\label{fig:xshooter}
\end{figure*}

The optical spectra from the two epochs at +371 days (VLT X-shooter) and +402 days (NTT EFOSC2) are shown in  Fig. \ref{fig:xshooter} (both dereddened and redshift corrected). To reduce noise in the X-shooter spectrum it was smoothed substantially using a Savitzky-Golay filter of second order, 31 points. Given the position of SN 2012ec in a bright region of the host galaxy, the slits included significant light from nearby stellar populations. The slope of the spectra and observed narrow absorption due to the Balmer series indicate significant contamination from OB-stars. To make a comparison of the observed spectrum with a SN model, we therefore need to add a model component for this background galaxy light. 

 

\begin{figure*}
\includegraphics[width=1\linewidth]{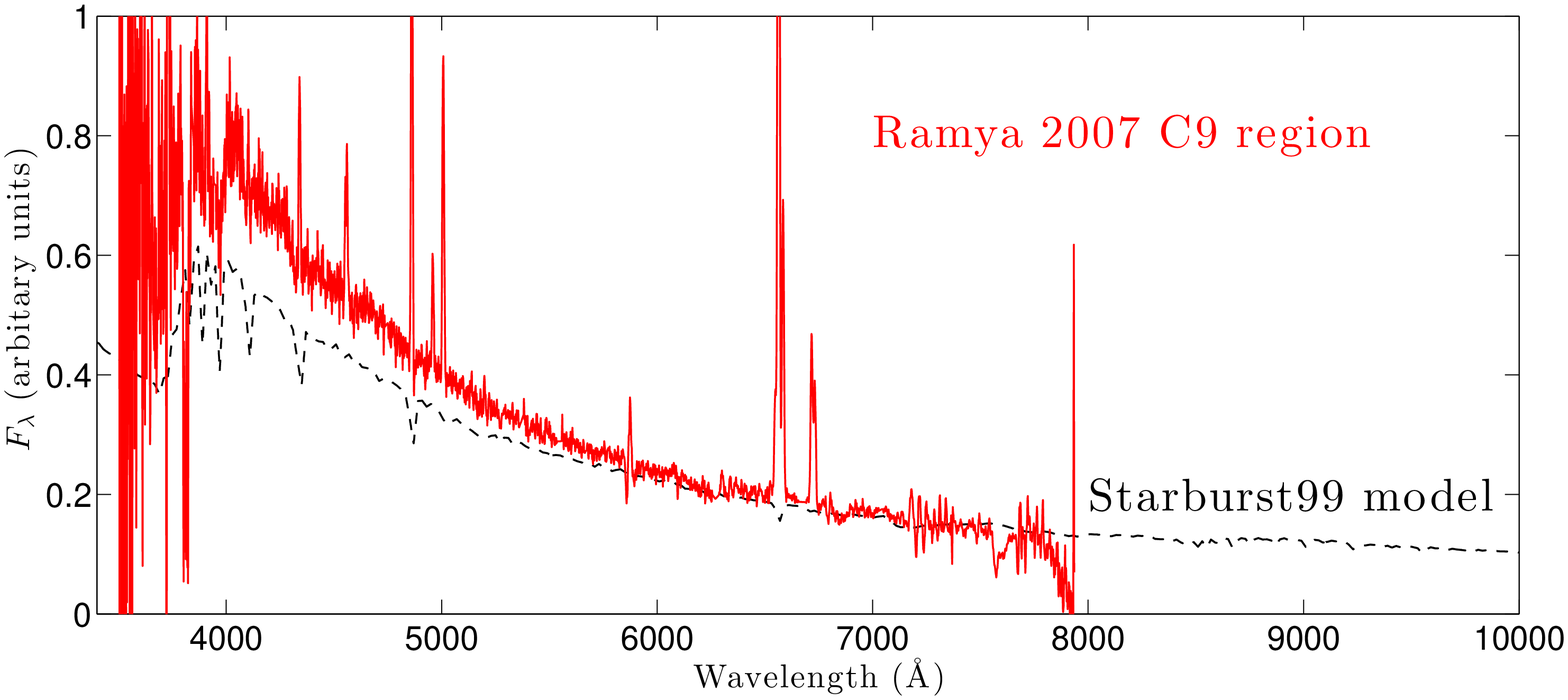} 
\caption{Starburst99 galaxy model (black dashed line), and a spectrum of the C9 region (in which SN 2012ec exploded) of NGC 1084, from \citet{Ramya2007}, dereddened with $E_{B-V}=0.45$ mag (red line).}
\label{fig:ramya}
\end{figure*}

To model the galactic background, we used Starburst99 \citep{Leitherer1999} to compute a model with a single starburst of standard parameters\footnote{Two-component IMF with $\alpha=1.3$ for $M=0.1-0.5$ \msun\ and $\alpha=2.3$ for $M=0.5-100$ \msun, Geneva 2012 tracks with no rotation, wind model ``Evolution'', atmosphere model ``Pauldrach/Hillier'', RSG microturbulence 3 \kms, and solar metallicity.}. The spectrum of this model at 30 Myrs is plotted as a black dashed line in Fig. \ref{fig:ramya}.  
Also plotted here (red line) is a spectrum of the star-forming region C9 in which SN 2012ec exploded, presented by \citet{Ramya2007}. The Ramya spectrum has been dereddened with $E_{\rm B-V}=0.45$ mag \citep[as estimated by][]{Ramya2007}. The comparison shows that the model spectrum is a reasonable representation of the galactic light around SN 2012ec, although the C9 region is a quite large region of the galaxy ($\sim$30\arcsec) and we lack knowledge of how much conditions change locally within this region (both intrinsic spectra and extinction).

Plotted in Fig. \ref{fig:xshooter}, as solid blue lines, are the sum of the galaxy model and the spectrum of the new 15 \msun~model computed with a \ni~mass of 0.03 \msun. The galaxy component is plotted as a black dashed line. The flux scale of the galaxy model is arbitrary and has been chosen to give overall good fits of the galaxy + SN model at each epoch. As the different epoch spectra have different slit widths, slit angles, and image qualities, the amount of galaxy light within the SN extraction window varies between them. In the +402d spectrum the flux (and thus our galaxy model scaling factor) is a factor 3.6 higher than for the +371d spectrum.

The sum of the galaxy and SN models fits the observed spectrum reasonably well at both epochs. The continuum slope is reproduced across a fairly wide wavelength range and the Balmer absorption lines and Balmer jump are reasonably well matched. The classic nebular emission lines seen in Type IIP SNe at these stages (see J12 and J14) are reproduced satisfactorily,  but there are two major and important discrepancies. The SN model overproduces the calcium lines ([Ca II] \wll7291, 7323 and the Ca II NIR triplet) and underproduces [Ni II] \wl7378. We investigate the formation of these lines further in Sect. \ref{sec:modellines}. 

\subsubsection{Oxygen lines and progenitor mass}
Using the method described in J12 to measure line luminosities, we measure the observed [O I] \wll6300, 6364 luminosity to $L=2.8\e{38}$ \ergs~at 371 days and $L = 2.9\e{38}$ \ergs~at 402 days.  The [O I] \wl5577 line is too weak and noisy for any meaningful luminosity measurements, which otherwise can provide a temperature constraint and a direct handle on the oxygen mass (see J14 for an example). 

The J12/J14 model luminosities of [O I] \wll6300, 6364 at 371 days for $M_{\rm ZAMS}=12, 15, 19, 25$ \msun~are $4.0, 8.0, 20, 21\e{38}$ \ergs (for a \ni~mass of 0.062 \msun). Assuming a direct proportionality with the \ni~mass \footnote{An approximation we can validate by comparing the new 15 \msun~model (M($^{56}$Ni) = 0.03 \msun) with the old  (M($^{56}$Ni) = 0.062\msun); the [O I] \wll6300, 6364 luminosity scales almost exactly with the $^{56}$Ni mass.}, these values for a $^{56}$Ni mass of 0.030\msun~are $1.9, 3.9, 9.5$, and $10\e{38}$ \ergs. The observed luminosity of $2.8\e{38}$ \ergs~thus corresponds to a progenitor ZAMS mass of $13-14$ \msun. 
At 402 days, the measured value of $2.9\e{38}$ \ergs~should be compared with the model values of $1.4, 2.9, 7.1$, and $7.7\e{38}$ \ergs, giving $M_{\rm ZAMS}=15$ \msun~as the best fit. We conclude that the nebular oxygen lines suggest a progenitor mass $M_{\rm ZAMS}=13-15$ \msun.


\subsubsection{Model lines in the 7100-7500 \AA\ region}
\label{sec:modellines}

We now aim to analyze in more detail the spectral region between 7100 - 7500 \AA, where the observed line strengths show the strongest discrepancies with the spectral model. Inspection of the models shows that lines from calcium, iron, and nickel are the only ones produced in any significant strength in this region. 

\paragraph*{Calcium lines}
The [Ca II] \wll7291, 7323 and Ca II NIR lines are weak in the observed spectra of SN 2012ec, both in comparison to the models as well as to most observed Type IIP SNe at this epoch. In the \citet{Maguire2012} sample of nine Type IIP SNe, the [Ca~II] \wl7291, 7323/[Fe II] \wl7155 ratio at 370 days spanned an interval $4-20$, whereas for SN 2012ec it is $\sim$1. Comparison with other emission lines gives a similar picture; the calcium lines are intrinsically weak in SN 2012ec.\\

As originally demonstrated by \citet{Li1993} \citep[see also][]{Kozma1998II}, the calcium lines originate as cooling emission from the hydrogen zone. We find the same conclusion with our models here; over 2/3 of the [Ca II] \wll7291, 7323 luminosity, and over 9/10 of the Ca II NIR luminosity, originate from hydrogen gas between 200-500 days. The rest of the [Ca II] \wll7291, 7323 luminosity has contributions from the Fe/He, Si/S, and He/C zones, all giving about 10\% each. The typical fraction of the total SN luminosity emerging in these forbidden calcium lines are 5-10\% in the models.

Being formed primarily as cooling of hydrogen-rich gas, their strength depends on the mass of the hydrogen zone, the heating rate per unit mass of this zone, and the fraction of cooling that is done by [Ca II]. Several different scenarios can thus lead to weak calcium lines, but some are more likely than others. A simple analysis is complicated by the fact that hydrogen zone material is present in many regions of the SN ejecta, both in the envelope and in the core (due to mixing).

If an unusually low hydrogen zone mass is responsible, we would expect to see unusually weak hydrogen lines as well. Both H$\alpha$ (Fig. \ref{fig:xshooter}) and Pa$\beta-\delta$ (Fig. \ref{fig:185d}) are quite strong and reasonably well reproduced by the model. Pa$\alpha$ looks weak but is compromised by its location in the middle of the telluric band. An unusually low ejecta mass was also not indicated by hydrodynamical modelling by Barbarino et al. (2014, submitted).

The heating rate per unit mass depends on the \ni~mass and the mixing of \ni~and hydrogen material. As the line luminosities to first order scale with the \ni~mass, the unusually low ratio of calcium lines to other lines in SN 2012ec cannot easily be understood as due to a low \ni~mass (which is also measured to be quite typical, 0.03 \msun). An unusually weak mixing of \ni~gas with hydrogen gas could possibly reduce the gamma-ray deposition in the H zone, but then again the hydrogen lines would be expected to decrease as well.

An low fraction of cooling done by Ca II could have three fundamental causes; a low calcium abundance compared to other metals, a low fraction of calcium in the Ca II state, or a temperature that does not favor Ca II emission. Starting with the last possibility, the models show that [Ca II] \wll 7291, 7323 is a strong coolant over the whole nebular evolution through a broad range of temperatures (J14). This suggests that an unusually low temperature is not the main cause of the weak lines.
The first possibility, an unusually low abundance of calcium relative to other metals, is also not a very plausible explanation as it would have to be lower by a factor 3-4 which seems unlikely.  A peculiarly low calcium abundance relative to the other elements is not a known feature of chemical abundances in starforming disk galaxies, and as the star-forming clouds in spirals will have been enriched in metals by a great number of stars and SNe any strong deviations from Initial Mass Function-weighted ratios is difficult to conceive of). 
We note that the metallicity itself of the progenitor will not affect the hydrogen zone cooling line strengths significantly. This is because cooling timescales are short so thermal equilibrium is established and all heating is instantaneously reemitted as cooling. Line cooling is the only efficient channel so lower metallicity will just lead to somewhat higher temperature so that the collisional pumping rates increase to compensate for the lower amounts of metals.

The second possibility, an unusual ionization balance of calcium, looks like a more plausible solution, because the ionization equilibrium in the models is close to switching between Ca II and Ca III as the dominant ion. At 370d the Ca II fraction is 25\% in the innermost H zone (75\% Ca III) and increases slowly outwards. Thus, the Ca II fraction is sensitive to small changes in physical conditions such as density and ionizing radiation field. It is quite plausible that these could vary with a factor of a few between different SNe. Dependency of the calcium ionization equilibrium on ejecta structure would be an interesting topic for further study. Other efficient coolants of the hydrogen gas around 400d are Mg II \wll2795, 2802, Fe II (several UV, optical, and NIR lines), and [O I] \wll6300, 6364. Hydrogen and helium usually heat the gas rather than cool it as recombinations and non-thermal excitations populate meta-stable (intrinsically meta-stable as well as effectively meta-stable through optical depth) states that are then collisionally deexcited.

We finally mention the possibility that the line at 7378 \AA~(which the model identifies as [Ni II] \wl7378) is actually a redshifted (3000 \kms) [Ca II] \wll7291, 7323. Several aspects safely rule out this option; the [Ca II] \wl7291, 7323 doublet is clearly seen close to its rest wavelength (it is just weak); no other lines in the spectra (including the H lines, which come from the same zone) show any detectable redshifts; the redshift would have to give exact wavelength coincidence with [Ni II] \wl7378 (which is the second strongest line in the model in this region); the Ca II NIR lines are as much weakened as the [Ca II] \wll7291, 7323 lines and show no redshift; and finally a strong [Ni II] 1.939 $\mu$m line matches the theoretical prediction following a [Ni II] \wl7378 identification (Sect. \ref{sec:nir}).


\paragraph*{Iron lines}
In the J14 models at 370 days, there are two distinct iron lines in the 7100-7500 \AA\ region; [Fe II] \wl7155 and [Fe II] \wl 7453. These lines arise from the same upper level (3d$^7$a$^2$G$_{9/2}$) with Einstein A-coefficients $A_{7155} = 0.146$ s$^{-1}$ and $A_{7453} = 0.0477$ s$^{-1}$, respectively. The lines are optically thin ($\tau_{7155}=0.08$ and $\tau_{7453}=0.02$), and so their luminosity ratio is $L_{7453}/L_{7155}= A_{7453}h\nu_{7453}/A_{7155}h\nu_{7155} = 0.31$. 

There is also some emissivity in [Fe II] \wl7172 (24\% of [Fe II] \wl7155) and [Fe II] \wl7388 (19\% of [Fe II] \wl7155). These lines arise from 3d$^7$a$^2$G$_{7/2}$, 0.07 eV above the 3d$^7$a$^2$G$_{9/2}$ state, and have $A_{7172} = 0.0551$ s$^{-1}$ and $A_{7388} = 0.0421$ s$^{-1}$; for any temperature they will thus be significantly weaker than [Fe II] \wl7155. Both 3d$^7$a$^2$G$_{9/2}$ and 3d$^7$a$^2$G$_{7/2}$ are close to Local Thermodynamic Equilibrium (LTE), with departure coefficients 0.83 and 0.85, respectively.

\paragraph*{Nickel lines}
The strongest nickel lines in this spectral region are [Ni II] \wl7378 and [Ni II] \wl7412. These lines come from 4s$^2$F$_{7/2}$ and 4s$^2$F$_{5/2}$ and have $A_{7378}=0.23$ s$^{-1}$ and $A_{7412} = 0.18$ s$^{-1}$. The lines are optically thin ($\tau_{7378}=0.04$ and $\tau_{7412} = 0.01$), and the parent states are close to LTE, with departure coefficients 0.94 and 0.92, respectively. The model ratio $L_{7412}/L_{7378} = 0.31$ at 370 days is therefore close to the theoretical value assuming LTE and optically thin conditions.

\paragraph*{Other lines}
Three additional lines are discernible in the models; He I \wl7281, Fe I \wl7207, and [Ni I] \wl7393. These lines are all quite weak, the [Ni I] \wl7393 line is for instance more than ten times weaker than [Ni II] \wl7378. 
We can conclude that the spectral region $7100-7500$ \AA\ is dominated exclusively by the components [Ca II] \wll7291, 7323, [Fe II] \wl7155, [Fe II] \wl7172, [Fe II] \wl7388, [Fe II] \wl7453, [Ni II] \wl7378, and [Ni II] \wl7412.

\subsubsection{Gaussian fits to the 7100-7500 \AA\ region}
\label{sec:gauss}
 
As shown in Fig.\,\ref{fig:xshooter}, the SN nebular model does not reproduce well the relative strengths of calcium, iron, and nickel lines in this spectral region. The [Ni II] \wl7378 line is particularly interesting because it provides a rare opportunity to determine the mass of stable nickel produced in the explosion. Instead of a full forward modelling approach, we in this section measure the observed line luminosities and apply analytical line formation equations to determine the element abundance ratios.  

\begin{figure*}
\includegraphics[width=1\linewidth]{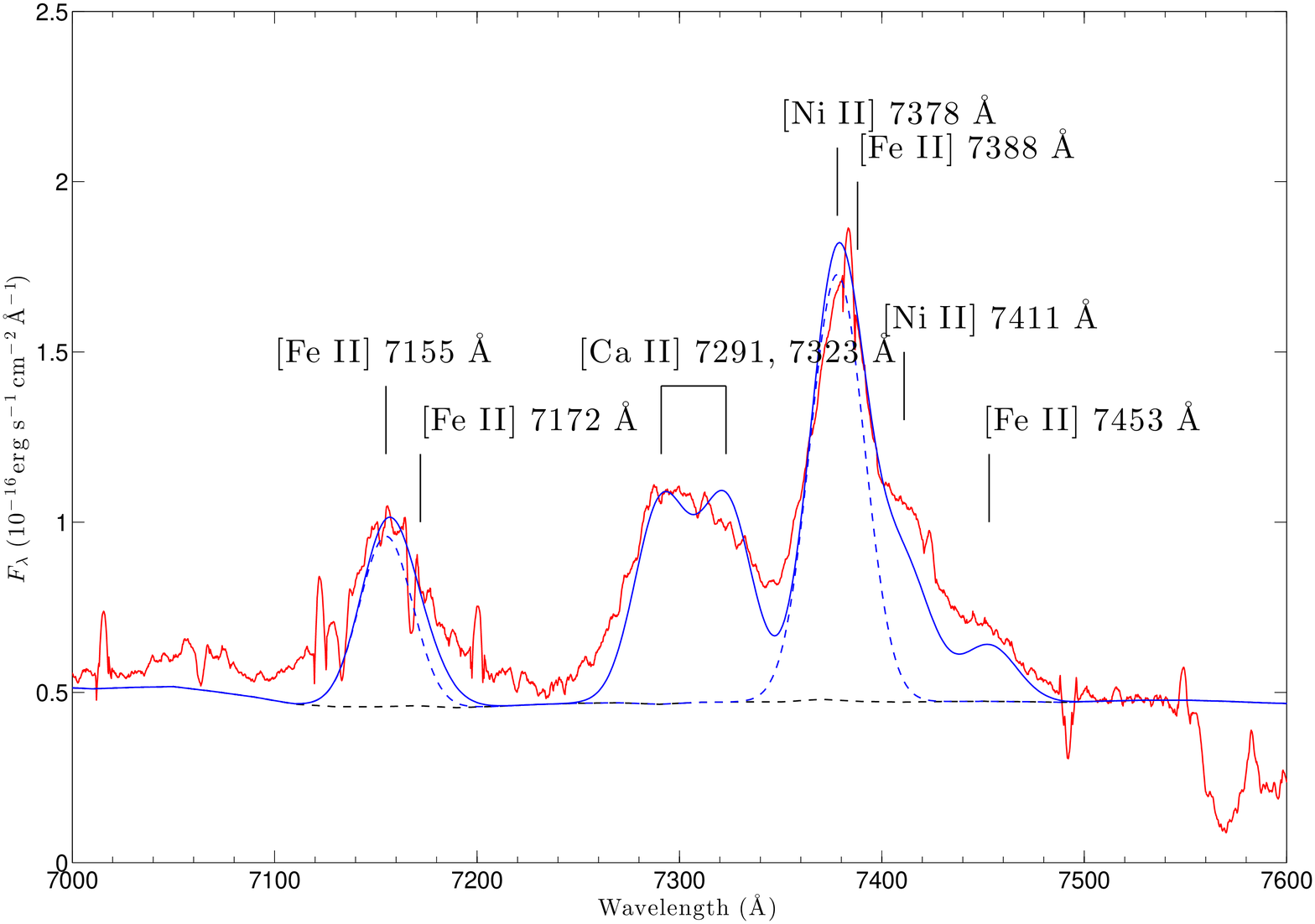} 
\caption{X-shooter spectrum between 7000-7600 \AA~(dereddened and redshift corrected, red), and the Gaussian fit described in Sect. \ref{sec:gauss} (blue). The components of the two lines that we use to derive the Ni/Fe ratio ([Fe II] \wl 7155 and [Ni II] \wl 7378) are plotted with dashed lines. Note that the [Fe II] \wl7172 and [Fe II] $\lambda$7388 components are constrained to be weak from atomic structure arguments. Hence the emission lines at 7155 and 7378 \AA\ are dominated by [Fe II] \wl7155 and [Ni II] \wl7378, respectively.}
\label{fig:gaussfit}
\end{figure*}

The model analysis in Sect. \ref{sec:modellines} showed that there are seven dominant line transitions present in the 7100-7500 \AA\ region, with constraints on some of the line ratios from the same species.
We construct a fit to this spectral region using Gaussian components of these lines ([Ca II] \wll7291, 7323, [Fe II] \wl7155, [Fe II] \wl7172, [Fe II] \wl7388, [Fe II] \wl7453, [Ni II] \wl7378, and [Ni II] \wl7412). We force the relative luminosities of lines from a given element to have the same ratios as in the model, so the iron lines are constrained by $L_{7453} = 0.31 L_{7155}$, $L_{7172}=0.24 L_{7155}$, $L_{7388} = 0.19 L_{7155}$, and the nickel lines are constrained by $L_{7412} = 0.31 L_{7378}$. We also use a single line width for all lines, the FWHM velocity $\Delta V$. The free parameters are then $L_{7291,7323}$, $L_{7155}$, $L_{7378}$, and $\Delta V$. We also add the galactic model component described in Sect. \ref{sec:galaxymodel} to the fit. As shown in Fig. \ref{fig:gaussfit}, a good fit is obtained for $L_{7291,7323} = 1.4\e{38}$ \ergs, $L_{7155}= 5.8\e{37}$ \ergs, $L_{7378} = 1.5\e{38}$ \ergs, and $\Delta V=1300$ \kms. From this we determine a ratio $L_{7378}/L_{7155} = 2.6$.


\subsubsection{The Ni II/Fe II ratio}
\label{sec:niiifeiiratio}
The identification of distinct [Fe II] \wl7155 and [Ni II] \wl7378 lines, and their measured luminosities, can constrain the iron and nickel content in SN 2012ec with some analytic treatment. 
Assuming LTE and optically thin emission (which are good approximations according to the model, Sect. \ref{sec:modellines}), the emissivity ratio of [Ni II] \wl7378 to [Fe II] \wl7155 is
\begin{flalign}
&\frac{L_{7378}}{L_{7155}} = \nonumber  \\
&= \frac{n_{\rm{NiII}} g_{4s2SF7/2}^{\rm Ni II} \exp{\left(-\frac{1.68~\mbox{eV}}{kT}\right)} Z_{\rm{NiII}}(T)^{-1} A_{7378} h \nu_{7378}}{n_{\rm FeII} g_{3d7a2G9/2}^{\rm Fe II} \exp{\left(-\frac{1.96~\mbox{eV}}{kT}\right)} Z_{\rm FeII}(T)^{-1} A_{7155}h\nu_{7155}}~,
\end{flalign} 
where $n_{\rm NiII}$ and $n_{\rm Fe II}$ are the number densities of Ni II and Fe II, $g$ are statistical weights, and $Z$ are partition functions. Computation of the partition functions $Z_{\rm Ni II}(T)$ and $Z_{\rm Fe II}(T)$ shows their ratio to vary little with temperature, staying between $0.24-0.26$ over the temperature range $2000-6000$ K, so we can to good accuracy take $Z_{\rm Ni II}/Z_{\rm Fe II}$ = 0.25. Using the atomic constants $g_{4s2SF7/2}^{\rm NiII} = 8$, $g_{3d7a2G9/2}^{\rm FeII}= 10$, $A_{7378} = 0.23$ s$^{-1}$, and $A_{7155} = 0.146$ s$^{-1}$, we then get
%

\begin{equation}
\frac{L_{7378}}{L_{7155}} = 4.9 \left(\frac{n_{\rm Ni II}}{n_{\rm Fe II}}\right) \exp{\left(\frac{0.28~\mbox{eV}}{kT}\right)}~.
\label{eq:LTElineratio2}
\end{equation} 
This line ratio is a  powerful diagnostic of the Ni II to Fe II ratio and in turn the Ni to Fe ratio. It is relatively insensitive to temperature due to the similar excitation energies of the upper levels. It is also quite insensitive to density because the transitions have similar critical densities; deviation from LTE has therefore a similar impact on both lines. Furthermore, since iron and nickel have similar ionization potentials, they have similar ionization balances and the Ni II/Fe II ratio is likely to be close to the Ni/Fe ratio. Our models confirm this, showing that the Ni II/Fe II ratio deviates from the Ni/Fe ratio by less than 5\% at 370 days.

\begin{figure}
\includegraphics[width=1\linewidth]{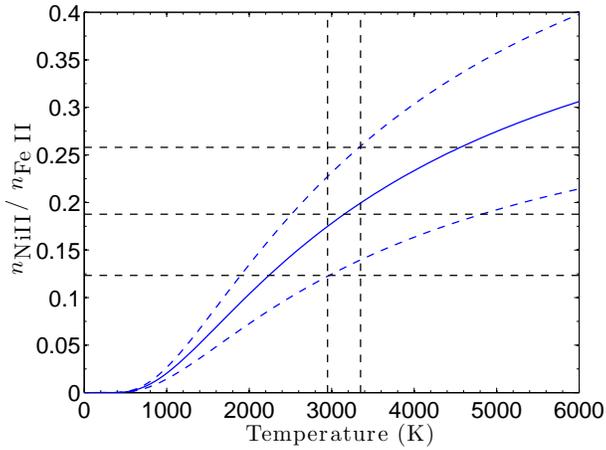} 
\caption{Derived Ni II/Fe II ratio as function of temperature (solid blue line shows the best estimate and the dashed blue lines give the error range). The temperature constraint from the [Fe II] \wl7155 luminosity (\protect\input{Tlow.txt}$ < T < $ \protect\input{Thigh.txt} K) gives a ratio $0.19\pm0.07$.}
\label{fig:massratiovsT}
\end{figure}

Using the measured line ratio of $L_{7378}/L_{7155}=2.6$, we can use Eq. \ref{eq:LTElineratio2} to compute the Ni II / Fe II ratio as function of temperature. This relationship is plotted in Fig. \ref{fig:massratiovsT} together with curves for $L_{7378}/L_{7155}$ ratios higher and lower by 30\% than 2.6 (our estimated error). The temperature constraints from the [Fe II] \wl7155 luminosity (\protect 2950$ < T < $ \protect 3350 K, see below) give a (number) abundance ratio of Ni II/Fe II $=0.19\pm0.07$, which we also then expect to be an accurate estimator of the Ni/Fe ratio. The solar abundance ratio is 0.06 (see Sect. \ref{sec:discussion}), and in the $M_{\rm ZAMS}=15$ \msun~model of \citet{Woosley2007}, the Ni/Fe ratio is 0.04. Thus, the reason that the SN model significantly underproduces the [Ni II] \wl7378 line in SN 2012ec is an abundance of stable nickel a factor $\sim$5 lower than needed.

\begin{figure}
\includegraphics[width=1\linewidth]{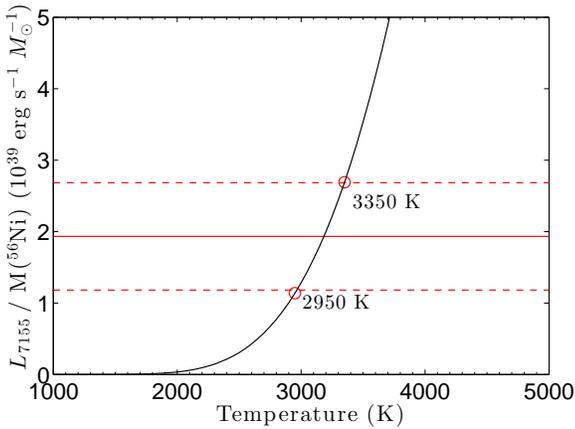} 
\caption{The measured value of $L_{7155}/M(\mbox{\ni})$ (red solid, errors as red dashed) compared to the theoretical value for LTE and optically thin emission as function of temperature (black line). To reproduce the observed value, the temperature is constrained to \protect\input{Tlow.txt}$ < T < $ \protect\input{Thigh.txt} K.}
\label{fig:Tconstr}
\end{figure}

\paragraph*{Constraints on temperature}
By 370 days all $^{56}$Ni has decayed and over 95\% of the daughter isotope $^{56}$Co has decayed to $^{56}$Fe. As this channel domininates the iron production, the iron mass is constrained by the measurement of the initial \ni~mass from the early nebular phase (Sect. \ref{sec:nimass}). Knowing the mass ($0.03\pm 0.01$ \msun), we can use the LTE expression for the [Fe II] \wl7155 luminosity to constrain the temperature. This assumes that most of the iron is in the form of Fe II. The fraction is 0.9 in our models at 370 days so this is likely a good approximation. It also assumes LTE. The departure coefficient is close to unity in the model (0.83), also validating this approximation. Figure \ref{fig:Tconstr} shows the measured value of $L_{7155}/ M(^{56}\mbox{Ni})$ compared to the theoretical value as a function of temperature, which is given by
\begin{equation}
\frac{L_{7155}}{M(^{56}\mbox{Ni})} = \frac{A_{7155} h\nu_{7155} g_{3d7a2G9/2}^{\rm Fe II} }{56 m_u Z_{\rm FeII}(T)}\exp{\left(\frac{-1.96~eV}{kT}\right)} ~.
\end{equation}
where $m_u$ is the atomic mass unit. The partition function is well described by $Z_{\rm FeII}(T) =  15 + 0.006\times T$ between $T=1000-5000$ K. The figure shows that the iron-zone temperature is constrained to \protect 2950$ < T < $ \protect 3350 K. This is in good agreement with the computed model temperature, which is $T=3180$\,K. 

\paragraph*{Contamination by primordial Fe and Ni}
In J12 and \citet{Maguire2012}, it was demonstrated that the total emission of primordial metals in the hydrogen zone is often comparable to the emission by newly synthesized metals in the core. We must therefore ask whether such emission may contribute to the [Fe II] \wl7155 and [Ni II] \wl7378 lines here.

At 370 days, the emission from primordial iron is about 40\% of the total [Fe II] \wl7155 emission. This contribution can be broken down into two parts, a contribution by hydrogen mixed into the core and having a similar velocity distribution as the synthesized iron, and the envelope part at higher velocities. These contributions give similar amounts of emission, about 20\% of the total each. The envelope component gives, however, a broad and flat-topped contribution to the line profile, which blends into the quasi-continuum as well as the [Ca II] \wll7291, 7323 doublet. Therefore it is does not become included in the luminosity we extract by fitting low-velocity Gaussians in Sect. \ref{sec:gauss}. On the other hand, the fit will contain the contribution from the in-mixed hydrogen zone, which is then of order 20\%/60\% $\sim$ 1/3 as strong as the synthesized iron component. 

Also [Ni II] \wl7378 has a similar contribution by primordial nickel ($\sim$40\%). A solar abundance of Ni to Fe (as is presumably present in the hydrogen zone), gives however a primordial [Ni II] \wl7378/[Fe II] \wl7155 ratio of only $0.7$. The contribution by primordial nickel to the measured [Ni~II] \wl7378 line luminosity (which is several times higher than the [Fe II] \wl7155 luminosity) must therefore be small ($\sim$$10\%$). The net effect of primordial contaminations is therefore that the measured Ni/Fe ratio may somewhat underestimate the Ni/Fe ratio in the iron zone, but not by more than about $\sim$1/3.

\subsection{Near-infrared}
\label{sec:nir}
\begin{figure*}
\includegraphics[width=1\linewidth]{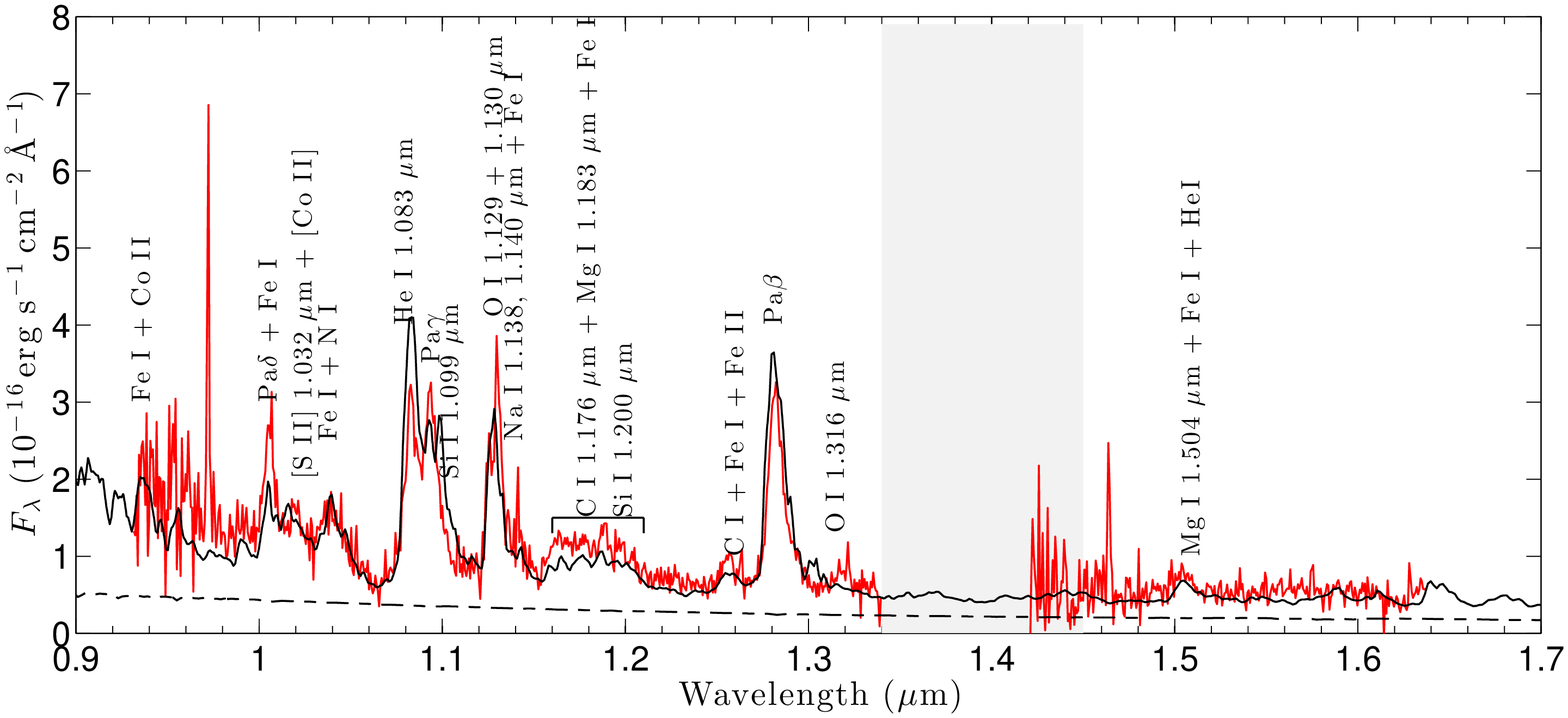}  
\includegraphics[width=1\linewidth]{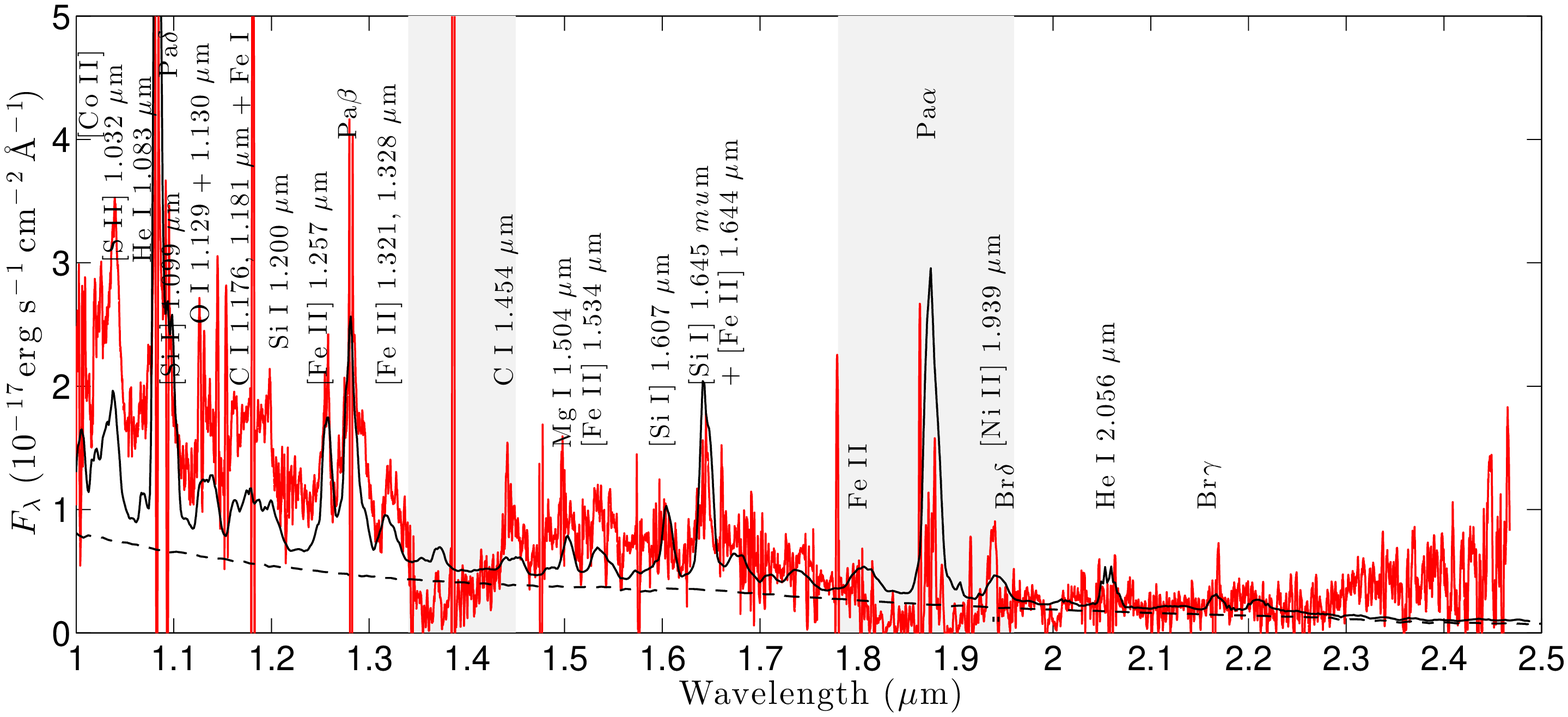}
\caption{{\bf Top} : SOFI near infrared spectrum of SN 2012ec (dereddened and redshift corrected) at +185d (red), and in black solid the sum of the galaxy model and the $M_{\rm ZAMS}=15$ \msun\ SN model of J14 at 212d scaled with 0.03/0.062 and a factor $\exp{\left(+27/111.4\right)}$. The galaxy model is shown as dashed black line. Line identifications are marked. {\bf Bottom} : X-shooter near infrared spectrum (dereddened and redshift corrected) at +371d (red), and in solid black the sum of the galaxy model and the $M_{\rm ZAMS}=15$ \msun~model with a \ni~mass of 0.03 \msun~computed here. The galaxy model is shown as a dashed black line. Telluric absorption bands are marked.}
\label{fig:185d}
\end{figure*}

Figure \ref{fig:185d} shows the observed NIR spectra at 185 days (top) and 371 days (bottom), compared with the galaxy model plus the 15 \msun~model of J14, scaled down with a factor $0.03/0.062$. The X-shooter spectrum has been smoothed to reduce noise, here with 61 points. The model provides identifications of the major emission lines, which are Pa$\beta$, Pa$\gamma$, Pa$\delta$, He I 1.083 $\mu$m, O I 1.129 + 1.130 $\mu$m, O I 1.316 $\mu$m, and several blends of C I, Mg I, Si I, S II, Fe I, Fe II, and Co II lines.

Of particular interest is the emission line at 1.94 $\mu$m. In the model there are only two lines with significant emission at this wavelength; [Ni II] 1.939 $\mu$m and Br$\delta$ 1.944 $\mu$m. The line is observed to be much stronger than Br$\gamma$ 2.166 $\mu$m, and is as narrow (FWHM $\approx$ 1300 \kms) as the iron-group lines between 7100-7500 \AA~(Sect. \ref{sec:gauss}). This feature was broader and weaker in the NIR spectrum at +306 days of SN 2012aw (J14), and the model matched the line strength well. Given the quantitative difference, we suggest that the feature is dominated by [Ni II] 1.939 $\mu$m in SN 2012ec.  With the caveat that the line is at the edge of the telluric band, we measure its luminosity to $L_{1.939~\mu m}=2.5\e{37}$ \ergs. 
[Ni II] 1.939 $\mu$m arises from the same upper level as [Ni~II] \wl 7378. Their optically thin line ratio is
\begin{equation}
\frac{L_{1.939~\mu m}}{L_{7378}} = \frac{A_{1.939}}{A_{7378}} \frac{h\nu_{1.939}}{h\nu_{7378}} = 0.14~,
\end{equation}
where we have used $A_{1.939} = 0.087$ s$^{-1}$ and $A_{7378} = 0.23$ s$^{-1}$. In SN 2012ec the measured ratio is
$\frac{L_{1.939~\mu m}}{L_{7378}} = 0.17$,
in excellent agreement. That the [Ni II] 1.939 $\mu$m line is close to the expected strength relative to [Ni II] \wl7378, and is again much stronger than in the model with Ni/Fe = 0.04 (the model line feature in Fig. \ref{fig:185d} is dominated by Br$\delta$), serves as an independent confirmation that the mass of stable nickel is high in SN 2012ec.

\section{Discussion and summary}
\label{sec:discussion}
The solar abundances of iron and nickel are well known, with close agreement between photospheric and meteoritic measurements. \citet{Lodders2003} report $\log \left(N_{Fe}/N_H\right)+12 = 7.48 \pm 0.03$ (meteoritic) and $\log \left(N_{Fe}/N_H\right)+12 = 7.45 \pm 0.08$ (photospheric) for iron, and $\log \left(N_{Ni}/N_H\right)+12 = 6.22 \pm 0.03$ (meteoritic) and $\log \left(N_{Ni}/N_H\right) + 12 = 6.22 \pm 0.13$ (photospheric) for nickel. The corresponding number ratio is Ni/Fe = 0.056 with an error of about 10\%, and the mass ratio is about 5\% higher (the mean atomic weights are 58.7 for nickel and 55.8 for iron), giving a mass abundance ratio of 0.059. \citet{Asplund2009} report similar values.
\citet{Hinkel2014} analyzed metal abundance variations in the solar neighborhood, reporting [Ni/Fe] = $-0.15_{-0.07}^{+0.15}$ dex at solar Fe/H metallicity, with similar values at other metallicities between [Fe/H] = -0.6 to +0.6 dex. This corresponds to Ni/Fe ratios of $0.6-1$ times the solar value; the spread in the Ni/Fe ratio throughout the local Universe is thus small.

Analysis of strong emission lines of [Ni II] \wl7378 and [Ni II] 1.939 $\mu$m in SN 2012ec indicates a strongly supersolar Ni/Fe production ratio of $0.20\pm0.07$ (by mass), or $\left(\mbox{Ni/Fe}\right)/\left(\mbox{Ni/Fe}\right)_\odot$ = $3.4\pm 1.2$. Estimates of the Ni/Fe ratio has been presented (to our knowledge) for four previous core-collapse SNe. For SN 1987A, \citet{Rank1988} reported a measurement of the [Ni II] 6.636 $\mu$m luminosity at 262 days. Assuming LTE, optically thin conditions, and that most nickel is singly ionized, this luminosity gave a Ni mass of $2\e{-3}$ \msun, which with $M(\mbox{Fe})=0.075$ \msun~from the \ni~measurement gives Ni/Fe = 0.027, less than half the solar value. The analysis is insensitive to temperature due to the low excitation potential (0.18 eV) for the first excited state giving the [Ni II] 6.636 $\mu$m line. 
\citet{Wooden1993} extended the analysis to later times, finding a similar value. In the ejecta models of J12/J14, LTE is valid for the [Ni II] 6.636 $\mu$m line throughout this period. Regarding ionization, we find about 80\% of the nickel being singly ionized between 250 and 450 days, with most of the remainder being Ni III. The most questionable assumption in the \citet{Rank1988} and \citet{Wooden1993} derivation is that of optically thin emission; in our models the [Ni II] 6.636 $\mu$m line has an optical depth of 3.2 at 250 days and 1.5 at 450 days. This optical depth leads to an underestimate of the Ni mass by a factor of 2-3 if one uses the optically thin formula. Taking the optical depth into account, the Ni/Fe ratio in SN 1987A is around solar or somewhat higher.  

\citet{Jerkstrand2012} calculated the model luminosity of [Ni II] 6.636 $\mu$m at 350d using the \citet{Woosley2007} nucleosynthesis models, finding good agreement with the observed line in SN 2004et \citep{Kotak2009}. As mentioned above, however, a caveat is that the nickel line has an optical depth of a few at these epochs, so the model luminosity depends not only on the nickel mass but also on the volume of the nickel-containing gas. This could fortunately be constrained from fits of various optically thick lines (see Figs. 11 and 12 in J12), so the fit of the [Ni II] 6.636 $\mu$m line is still constraining for the nickel mass in SN 2004et, giving a Ni/Fe ratio around solar or somewhat lower.

The [Ni II] \wl7378 line is one of the strongest lines emitted by the Crab nebula, with the Ni/Fe ratio estimated to be $60-75$ times the solar value \citep{Macalpine1989, Macalpine2007}. The variation with spatial position is relatively small, and imaging shows the nickel and iron lines to arise from the same filaments \citep{Macalpine1989, Hester1990}. This extreme value is of interest in relation to the proposed scenario of an electron capture SN \citep{Nomoto1982}, which have been shown to produce very large Ni/Fe ratios \citep{Wanajo2009}. 

A detection of [Ni II] \wl7378 was reported also for the broad-lined Type Ic SN 2006aj \citep{Maeda2007,Mazzali2007} which was associated with an X-ray flash \citep{Mazzali2006}. The [Ni II] identification was somewhat hampered by high noise levels combined with very large widths of the emission lines. The nebular spectra were modelled to estimate a mass of stable nickel of $0.02-0.05$ \msun, which with $M(^{56}\mbox{Ni})=0.2$ \msun~from the light curve gives Ni/Fe = $0.10-0.25$, similar to the value derived here for SN 2012ec.

\begin{figure*}
\includegraphics[width=0.49\linewidth]{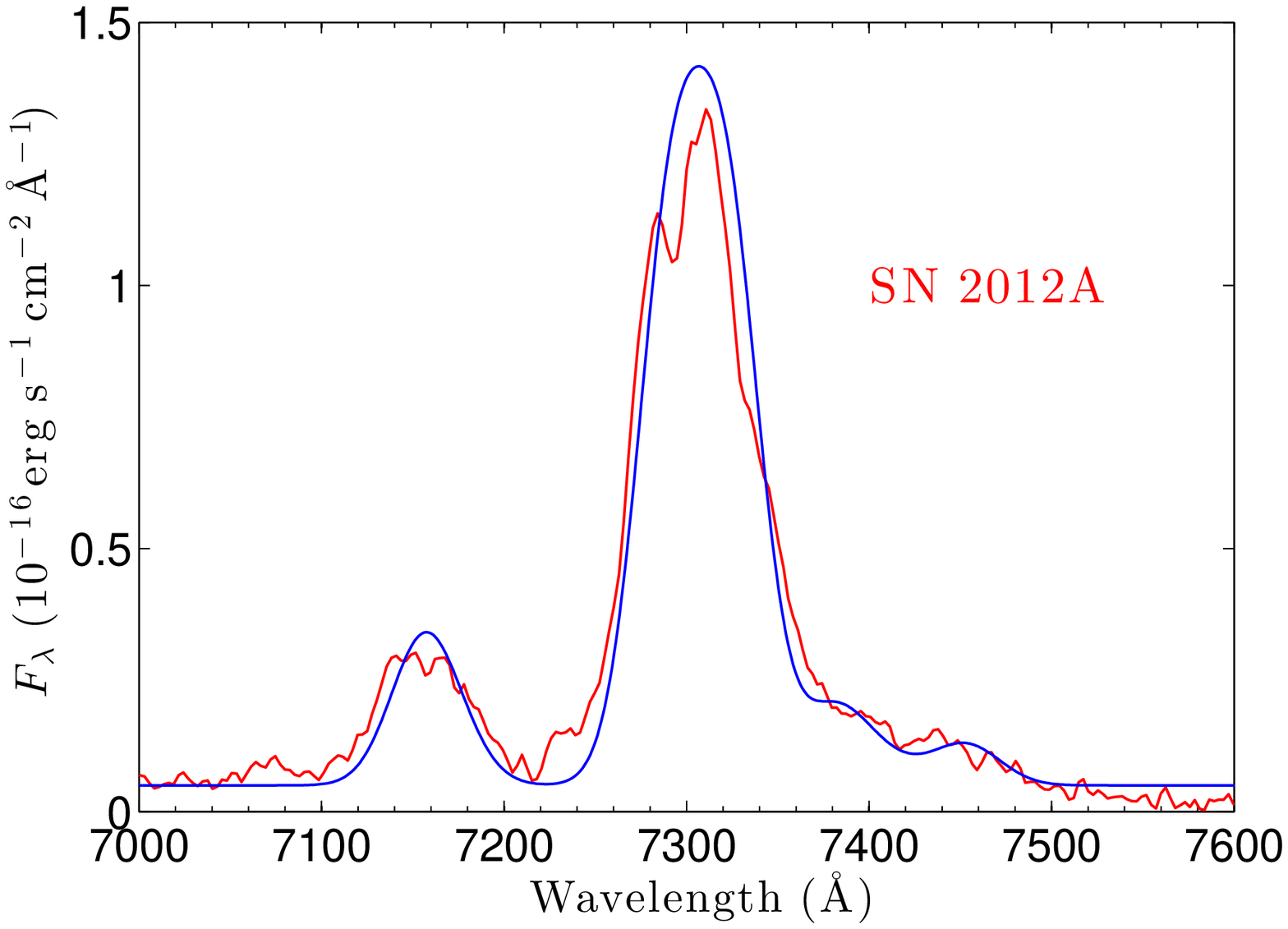} 
\includegraphics[width=0.49\linewidth]{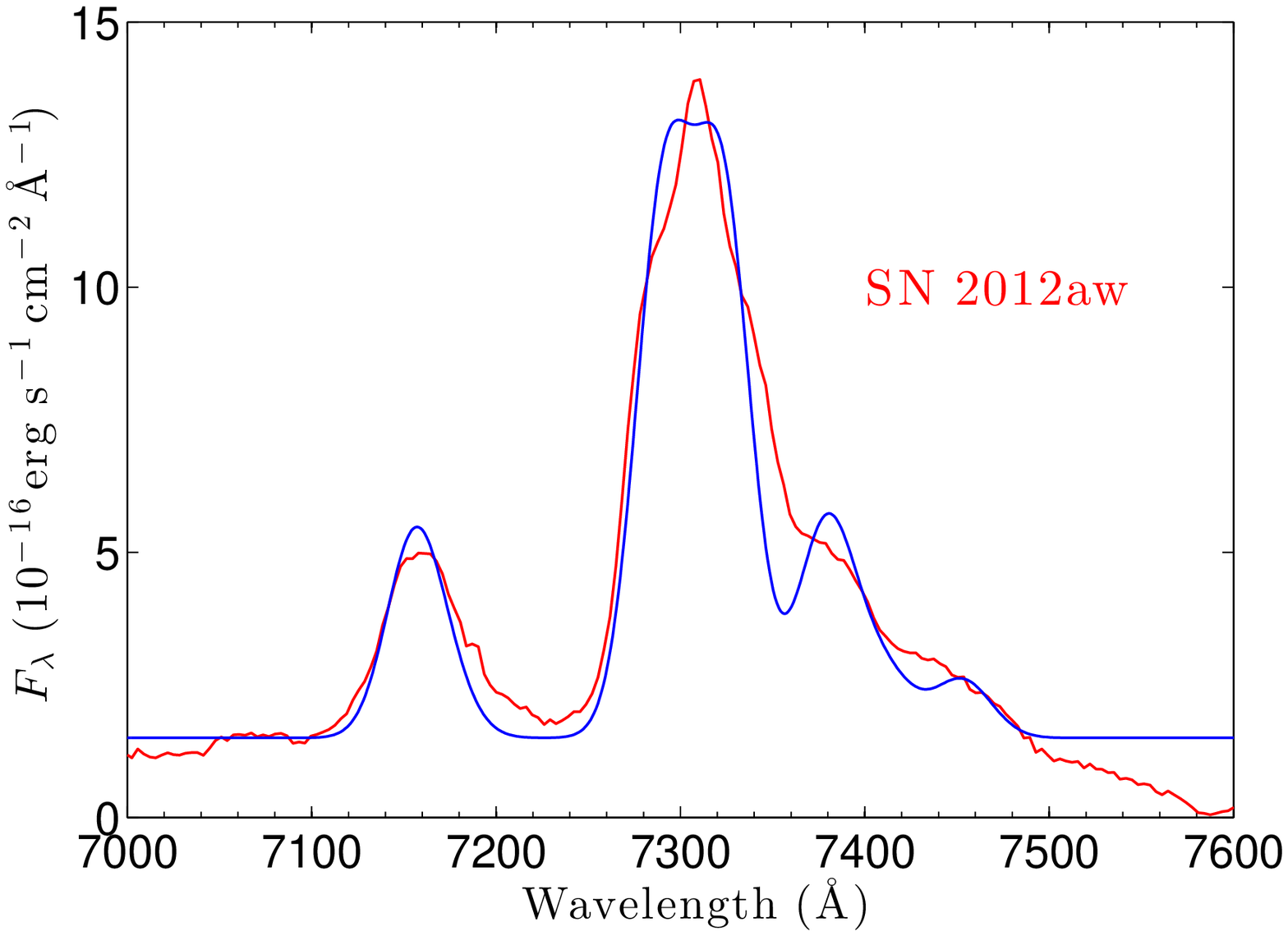} 
\caption{Spectra of SN 2012A at +393d (left) and SN 2012aw at +369d (right), and Gaussian fits following the same formalism as for the SN 2012ec fits. For SN 2012A the fit values are $\Delta V=1700$ \kms, $L_{7291,7323}= 1.1\e{38}$ \ergs, $L_{7155}=1.3\e{37}$ \ergs, and $L_{7378}=5.3\e{36}$ \ergs. For SN 2012aw, the fit values are $\Delta V=1500$ \kms, $L_{7291,7323}= 9.0\e{38}$ \ergs, $L_{7155}=1.5\e{38}$ \ergs, and $L_{7378}=1.6\e{38}$ \ergs.}
\label{fig:twomore}
\end{figure*}

We repeated the analysis of SN 2012ec for two other recently observed Type IIP SN; SN 2012A and SN 2012aw. The 7000-7600 \AA~spectral region of these SNe around +400d (data from \citet{Tomasella2013} and J14), and Gaussian fits, are presented in Fig \ref{fig:twomore}. The [Ca II] \wll7291, 7323 lines are strong in both of these SNe, and [Ni II] \wl7378 is not distinct. With our Gaussian fits, we can still obtain estimates for the nickel line luminosities. We obtain $L_{7378}/L_{7155} \approx 0.4$ for SN 2012A and $L_{7378}/L_{7155} \approx 1.0$ for SN 2012aw. With the same analysis as for SN 2012ec (using the distances, extinction values, and \ni~masses from \citet{Tomasella2013} and J14), this translates to Ni/Fe ratios of $\sim 0.5$ times solar in SN 2012A and $\sim 1.5$ times solar in SN 2012aw. We warn, however, that the \ni~mass in SN 2012A is only $\sim$0.01 \msun \citep{Tomasella2013} so the physical approximations used (validated by models with \ni~masses of 0.062 and 0.03 \msun) may not be as good at this low \ni~mass. The \ni~mass in SN 2012aw is around 0.06 \msun, so that problem is not present for this SN. A roughly solar Ni/Fe ratio in SN 2012aw is supported by the good fit to the [Ni II] 1.939 $\mu$m + Br$\delta$ blend with the standard models in J14.

A summary of all reported measurements (that we are aware of) are presented in Table \ref{table:NiFemeas}. Of the seven measurements, three (the Crab, SN 2006aj, and SN 2012ec) show significantly supersolar production of Ni/Fe, which should put strong constraints on the progenitor structure and explosion dynamics of these SNe. The Ni/Fe production is for many types of explosions dominated by the \nif~/\ni~production. This ratio in turn depends on the neutron excess of the fuel, as well as the thermodynamic conditions for the burning \citep{Woosley1973}. Higher neutron excess favours the production of neutron-rich isotopes such as \nif, as does high entropy explosions which lead to large abundances of neutrons and $\alpha$-particles. In Jerkstrand et al. (in prep.) we investigate the type of explosive silicon burning that can produce a Ni/Fe ratio as high as in SN 2012ec. 

\begin{figure}
\includegraphics[width=1\linewidth]{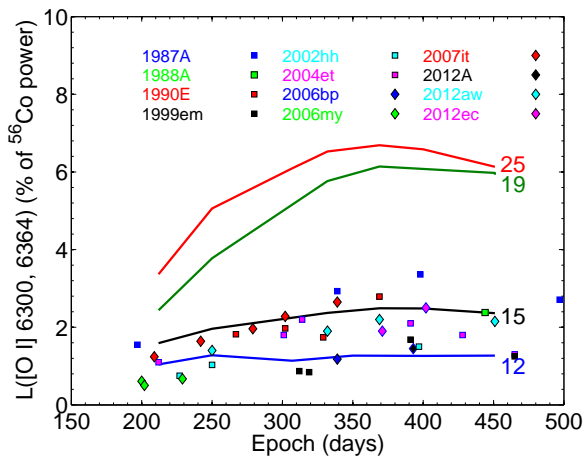} 
\caption{The [O I] \wll6300, 6364 luminosities (normalized to the $^{56}$Co~decay power) for twelve Type IIP SNe, compared to the J14 model tracks for different $M_{\rm ZAMS}$.}
\label{fig:oitracks}
\end{figure}

We have also analyzed the nebular [O I] \wll6300, 6364 lines, which match models with $M_{\rm ZAMS} = 13-15$ \msun. The progenitor analysis by \citet{Maund2013} found $M_{\rm ZAMS}=14-22$ \msun, and hydrodynamical modelling by Barbarino et al. (2014, submitted) favoured an ejecta mass of $\sim$13 \msun~(to be compared with total ejecta masses for $M_{\rm ZAMS} = 12, 15$, and 19 \msun~stars which are 9, 11, and 14 \msun~ in the \citet{Woosley2007} models). Within the errors, the methods seem to agree on an intermediate-mass progenitor. Figure \ref{fig:oitracks} shows the J14 model tracks for [O I] \wll6300, 6364 and the measured values for SN 2012ec and eleven other Type IIP SNe with high-quality nebular-phase spectra; SN 1987A\footnote{SN 1987A is strictly speaking a Type IIpec rather than a Type IIP, but we include it here because its nebular properties are expected to be similar to those of Type IIP SNe.} \citep{Phillips1990}, SN 1988A \citep{Turatto1993}, SN 1990 \citep{Benetti1994, Gomez2000}, SN 1999em \citep{Elmhamdi2003}, SN 2002hh \citep{Pozzo2006}, SN 2004et \citep{Sahu2006}, SN 2006bp \citep{Quimby2007}, 2006my \citep{Maguire2010}, SN 2007it \citep{Andrews2011}, SN 2012A \citep{Tomasella2013}, and SN 2012aw (J14). Values for distances, extinctions, explosion epochs, and \ni~masses were taken as derived in each data paper, or in their references \footnote{For SN 1988A, SN 2006bp, and SN 2006my no $^{56}$Ni masses have been presented, for these we used 0.030, 0.038 and 0.031 \msun~based on calibration to SN 1987A flux levels (assuming distances of 17, 15, and 22 MPc, respectively).}. All of these show O I lines suggesting $M_{\rm ZAMS}\lesssim 17$ \msun~progenitors; a full statistical analysis with more data is underway. 

SN 2012ec also shows an unusually weak [Ca II] \wll7291, 7323 feature (which indeed allowed the clear measurement of the [Ni II] \wl7378 line). According to the models, the [Ca II] \wll7291, 7323 doublet (as well as Ca II NIR triplet) originates mainly from the hydrogen zone in the supernova. A plausible explanation for variation in this line between different supernovae is varying ionization balance between Ca II and Ca III. However, other possibilities exist as well, and a more detailed study of Ca II line formation in Type IIP SNe would be desirable.

\begin{table*}
\caption{Measurements of the Ni/Fe ratio in core-collapse SNe.}
\begin{tabular}{cccccccccc}
\hline
SN           & Ni/Fe (times solar)   & Reference \\
\hline
Crab         & $60-75$    & \citet{Macalpine1989, Macalpine2007} \\
SN 1987A     & $0.5-1.5$  & \citet{Rank1988, Wooden1993}; this work\\
SN 2004et    & $\sim$1    & \citet{Jerkstrand2012}\\ 
SN 2006aj    & $2-5$      & \citet{Maeda2007, Mazzali2007}\\
SN 2012A     & $\sim 0.5$ & This work\\
SN 2012aw    & $\sim 1.5$ & This work\\
SN 2012ec    & $2.2-4.6$  & This work\\
\hline 
\end{tabular}
\label{table:NiFemeas}
\end{table*}

\section*{Acknowledgments}
We thank S. Sim, B. M\"uller, K. Nomoto, and R. Fesen for discussion, and R. Sethuram for contributing the NGC 1084 galaxy spectrum. We thank R. Kotak for use of LT data, P. Meikle for providing spectra of SN 2002hh and J. Andrews for providing spectra on SN 2007it. We have made use of the SUSPECT and WISEREP \citep{Yaron2012} databases. This work is based in part on observations collected at the European Organisation for Astronomical Research in the Southern Hemisphere, Chile as part of PESSTO, (the Public ESO Spectroscopic Survey for Transient Objects) ESO program 188.D-3003. The research leading to these results has received funding from the European Research Council under the European Union's Seventh Framework Programme (FP7/2007-2013)/ERC Grant agreements n$^{\rm o}$ [291222] (SJS) and n$^{\rm o}$ [320360] and STFC grants ST/I001123/1 and ST/L000709/1 (SJS).

\bibliographystyle{mn2e3}
\bibliography{bibl_part1}

\label{lastpage}
\end{document}